\documentclass[twocolumn,prl,showpacs]{revtex4}
\usepackage{amsmath,amssymb}
\usepackage[dvips]{graphicx,color}
\usepackage{bm}

\begin{document}

\title{
Current induced transition of anisotropic quantum Hall states
}

\author{Kazumi Tsuda}
\author{Nobuki Maeda}
\author{Kenzo Ishikawa}
\affiliation{Department of Physics, Hokkaido University, Sapporo
060-0810, Japan}
\date{\today}

\begin{abstract} 
We compare the energies of the striped Hall state and the anisotropic 
charge density wave (ACDW) state at half-filled third and higher Landau
levels in the system with injected currents. 
With no injected current, the ACDW state has a lower energy. 
We find that 
the striped Hall state becomes the lower energy state 
when the injected current exceeds a critical value. 
The critical value is estimated as about $0.04$-$0.05$ nA. 
\end{abstract}

\maketitle
One of the most fascinating states in the quantum Hall system 
is a highly anisotropic state. 
The highly anisotropic state has been found 
around half-filled third and higher
Landau levels (LLs) in ultra-high mobility samples 
at low temperature \cite{stripe_ex1,stripe_ex2}. 
The state reveals highly anisotropy in the longitudinal resistivity 
$\rho_{xx}$ and $\rho_{yy}$. 
In the experiments at the filling factor $\nu=9/2$, 
$\rho_{xx}$ is about $1000\Omega$ while $\rho_{yy}$ is about several
$\Omega$ at low temparetures of order tens of mK.  
In this state, no quantized plateaus in the Hall resistivity $\rho_{xy}$
have been found. 
While several states have been proposed to explain the experimental
results so far 
\cite{stripe0,stripe01,stripe02,stripe03,stripe04,stripe05,stripe06,
ACDW1,stripe07,stripe08,stripe09,stripe10,stripe1,stripe2}, 
it is still an open problem which state is realized in the experiments. 

\begin{figure}
 \includegraphics[width=9cm]{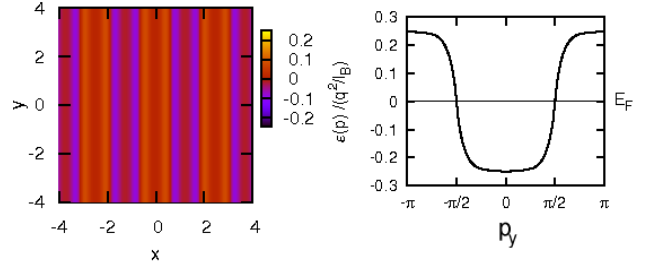}
 \caption{\label{fig:dens_ene_stripe}Left: Density modulation of 
the striped Hall state at the half-filled third LL. 
The uniform part is subtracted. 
The density is periodic in the $x$-direction and uniform in the
$y$-direction (stripe direction). 
The injected current flows easily in the stripe direction. 
Right: Energy spectrum 
of the striped Hall state at the same filling. 
It is uniform in the $p_x$-direction. 
The BZ is $|p_i|<\pi$ and 
$E_F$ is a Fermi energy. 
}
\end{figure}

\begin{figure}
 \includegraphics[width=9cm]{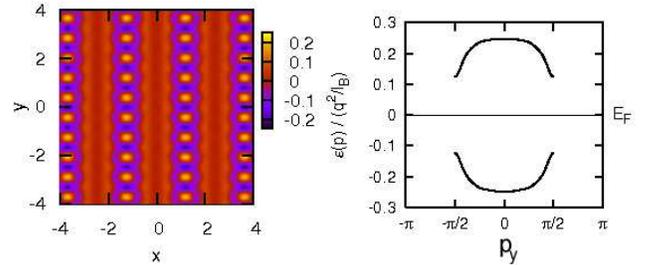}
 \caption{\label{fig:dens_ene_ACDW}Left: Density modulation of the ACDW
state at the half-filled third LL. The uniform part is subtracted. 
The density is modulated along stripes. 
Right: Energy spectrum of the ACDW state at the same filling. 
It depends on $p_x$ weakly and the spectrum at $p_x=0$ is plotted. 
The BZ is reduced to $|p_y|<\pi/2$ due to the modulations along stripes
and the two bands are formed. 
}
\end{figure}

\begin{figure}
 \includegraphics[width=4cm]{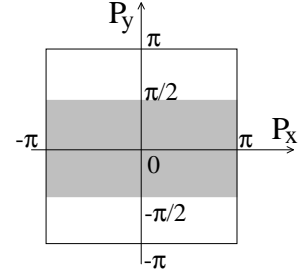}
 \caption{\label{fig:FS}
Fermi sea of the striped Hall state at half-filling 
in the case that the stripe direction faces the $y$-direction. 
The occupied state is represented by the dark region. 
The Fermi sea has the inter-LL energy gap in the
$p_x$-direction and is gapless in the $p_y$-direction. }
\end{figure}

We focus on two Hartree-Fock (HF) states among them. 
One is a striped Hall state and another is an anisotropic charge
density wave (ACDW) state. 
The striped Hall state is a unidirectional charge density wave state 
\cite{stripe0} 
and is a gapless state with an anisotropic Fermi surface
\cite{stripe1,stripe2}(Fig. \ref{fig:dens_ene_stripe}, Fig. \ref{fig:FS}). 
The ACDW state \cite{ACDW1} 
is a state which has the similar CDW order
perpendicular to the stripe direction as the striped Hall state, but 
in addition has the density-wave modulation along stripes (Fig. \ref{fig:dens_ene_ACDW}). 
The density modulation along stripes 
results in an energy gap. 
These properties suggest 
that the anisotropic state is the striped Hall state 
since 
the anisotropic longitudinal 
resistivity and the un-quantized Hall resistivity 
are naturally explained by the 
anisotropic Fermi surface \cite{stripe1,stripe2}, 
while it is difficult to explain these experimental features with the ACDW state
because of the energy gap. 
However it has been pointed out that the striped Hall state is unstable 
within the HF approximation to formation of modulations along 
stripes so that the ACDW state is the lower energy state \cite{ACDW1}. 
This has been 
an enigma 
as to the anisotropic states. 

In this paper, we study 
energy corrections due to injected currents flowing in the stripe
direction for the two HF states. 
The effect of the injected current has not been taken into 
account in the previous studies. 
The density modulations along stripes shown in
Fig. \ref{fig:dens_ene_ACDW} suggest that 
the effect of the injected current becomes larger in the ACDW state. 
When the current is injected, 
charges accumulate around both edges 
perpendicular to the current flow with the opposite sign 
as in the case of the classical Hall effect and 
the accumulated charges cause the energy enhancement via the 
Coulomb interaction. 
We calculate this type of energy corrections for the present HF states 
and find that the energy correction for the ACDW state is larger than
that for the striped Hall state. 

Let us consider the two-dimensional (2D) 
electron system in the $x$-$y$ plane with an
external 
magnetic field $B$ 
in the $z$-direction. 
The spin degree of freedom is ignored and 
the natural unit ($\hbar=c=1$) is used in this paper. 
In the von Neumann lattice (vNL) formalism \cite{vNL2,vNL3} which is
suitable for studying spatially periodic states, 
the present system is represented as a 2D lattice system and 
a momentum 
is defined in the Brillouin zone (BZ). 
We use the rectangular vNL base with a vNL asymmetry parameter $r_s$ in
the following calculation. 
The electron field operator is expanded by the vNL base as 
\begin{equation}
 \Psi({\bf x})=\int_{\rm BZ}\frac{d^2p}{(2\pi)^2}\sum_{l=0}^{\infty} 
  b_l({\bf p})\langle {\bf x}| f_l \otimes \beta_{\bf p}\rangle, 
\end{equation}
where $l$ denotes the LL index, $b_l({\bf p})$ is the anti-commuting 
annihilation operator with momentum {\bf p} defined in the BZ 
($|p_i|<\pi$), and 
$\langle {\bf x}| f_l \otimes \beta_{\bf p}\rangle$ 
is given in Ref. \onlinecite{vNL3}. 
The momentum state is the Fourier transform of the Wannier basis of the
vNL which is localized at ${\bf x}=a(r_s m, n/r_s)$, where $m$, $n$ are
integers and $a=\sqrt{2\pi/eB}$ is a vNL constant. 
We set $a=1$ unless otherwise stated. 
Note that the number of vNL unit cells 
is equal to the number of states within one LL. 
In the $l$th LL Hilbert space, the kinetic term is quenched and 
the Hamiltonian is given by the projected Coulomb interaction 
\begin{gather} 
 \mathcal{V}^{(l)}=\frac{1}{2} \int \frac{d^2 k}{(2\pi )^2} 
  : \rho_l({\bf k}) V({\bf k}) \rho_l(-{\bf k}):\notag,\\
 V({\bf k})=\frac{2\pi q^2}{k}\ (k\neq 0, q^2=\frac{e^2}{4\pi\epsilon}),
 \quad V(0)=0, 
 \label{eq:projected_Coulomb}
\end{gather}
where the colons represent normal ordering with respect to 
creation and annihilation operators, 
$\epsilon$ is the dielectric constant, and 
$\rho_l({\bf k})$ is a projected density operator. 
$\rho_l({\bf k})$ is represented by the vNL base as 
$\rho_l({\bf k})=F_l({\bf k})\bar{\rho}_l({\bf k})$
where $F_l({\bf k})=e^{-k^2/8\pi}L^0_l(k^2/4\pi)$ 
($L^0_l$ is a Laguerre polynomial)
and 
\begin{equation}
 \bar{\rho}_l({\bf k})= \int_{{\rm BZ}}\frac{d^2 p}{(2\pi)^2} 
 b^\dag_l({\bf p}) b_l({\bf p}-\hat{{\bf k}})
 e^{-(i/4\pi)\hat{k}_x(2p_y-\hat{k}_y)}. 
\end{equation}
Here $\hat{\bf k}=(r_sk_x,k_y/r_s)$.

{\it Striped Hall state.}--- 
The Hamiltonian (\ref{eq:projected_Coulomb}) 
has been diagonalized self-consistently in the HF
approximation for the striped 
Hall state and the ACDW state. 
We concentrate on the case of the filling factor $\nu=l+\nu^\ast$ with
$\nu^\ast=1/2$. 
The striped Hall state is obtained by assuming 
\begin{equation}
 \label{eq:stripe_density}
 \langle \bar{\rho}_l({\bf k}) \rangle _{\rm stripe}=
  \sum_{N_x}\Delta_l(N_x)(2\pi)^2
  \delta(k_x+\frac{2\pi N_x}{r_0})\delta(k_y),  
\end{equation}
where $N_x$ is an integer, 
$r_0$ is the period of the density in the $x$-direction, 
$\Delta_l(N_x)$ is an order parameter
determined self-consistently, 
and $\Delta_l(0)=\nu^\ast$. 
Note that the density (\ref{eq:stripe_density}) is uniform in the
$y$-direction in the coordinate space as shown in
Fig. \ref{fig:dens_ene_stripe}.  
The HF Hamiltonian is diagonalized
self-consistently by taking $r_s=r_0$. 
$\Delta_l(N_x)=(-1)^{N_x}\sin(\nu^\ast\pi N_x)/\pi N_x$ is 
a self-consistent solution and   
the resulting state has the Fermi sea, 
$|p_x|<\pi$ and $|p_y| < \pi\nu^\ast$ which gives the anisotropic Fermi
surface (Fig. \ref{fig:FS}) \cite{stripe1,stripe2}. 
Since the anisotropic Fermi surface has the inter-LL energy
gap in the $p_x$-direction and is gapless in the $p_y$-direction, 
the striped Hall state would have the anisotropic longitudinal
resistivity. The total energy depends on $r_s$ and 
the minimum energy per particle is given by $-0.3074$ ($r_s=2.474$) for
$l=2$ and $-0.2800$ ($r_s=2.875$) for $l=3$ in units of
$q^2/l_{\rm B}$, where $l_{\rm B}=\sqrt{1/eB}$ is the magnetic length.

{\it ACDW state.}---
The ACDW state is obtained by assuming \cite{ACDW0}
\begin{equation}
 \langle \bar{\rho}_l({\bf k}) \rangle_{\rm ACDW}=
  \sum_{\bf N}
  \Delta_l({\bf Q}_N)(2\pi)^2\delta^2({\bf k}-{\bf Q}_N), 
  \label{eq:ACDW_density}
\end{equation}
where ${\bf Q}_N=(2\pi N_x/r_{0x}, 2\pi N_y/r_{0y})$, 
$r_{0x}$ and $r_{0y}$ are the periods of the density in the $x$-direction
and the $y$-direction respectively, 
$\Delta_l({\bf Q}_N)$ is an order parameter determined
self-consistently, 
and $\Delta_l(0)=\nu^\ast$. 
The density modulations in both directions generate the energy bands 
and the BZ is reduced to a smaller BZ as shown in Fig. \ref{fig:dens_ene_ACDW}. 
The direction with a shorter period of the density modulation is
referred to as {\it stripe direction} of the ACDW state in this paper. 
We concentrate on the case of $\nu^\ast=1/2$. 
The HF Hamiltonian is diagonalized under the assumption
(\ref{eq:ACDW_density}) by taking
$r_s=r_0$. 
Since the number of ACDW unit cells is equal to the number of
electrons within one LL, the area of the ACDW unit cell is just twice as
large as the vNL unit cell. 
Hence the self-consistent solution gives two energy bands and the BZ is 
reduced to the half size of the original BZ, $|p_x|<\pi$ and $|p_y|<\pi/2$. 
The total energy depends on $r_s$ and the minimum energy per particle is
given by $-0.3097$ ($r_s=0.82$ or $2.44$) for $l=2$ and $-0.2814$
($r_s=0.70$ or $2.86$) for $l=3$ in units of $q^2/l_{\rm B}$, where 
there are two values of $r_s$ at each LL due to the
$\pi/2$-rotational symmetry. 
The magnitudes of energy gaps are $0.2470$ for $l=2$ and $0.1967$ for
$l=3$ in units of $q^2/l_{\rm B}$. These values are estimated as of
order 10K for a few tesla in terms of temperature while 
experiments for the anisotropic states have shown 
the anisotropic longitudinal resistivities at 
much lower temperatures than the energy gap of order tens of mK. 
Therefore it is difficult to explain the experiments
with the ACDW state. However the energy of the ACDW state is slightly
lower than that of the striped Hall state at each LL  
(the energy differences $\Delta E_0$ are $2.3\times10^{-3}$ for $l=2$ and 
$1.4\times 10^{-3}$ for $l=3$ in units of $q^2/l_{\rm B}$). 
This was one of the remained issues for the anisotropic states. 

{\it Energy corrections due to current.}---
Next we take into account the effect of the injected current on the
present two HF states and calculate energy corrections due to 
current. 
The lengths of the 2D electron system in the $x$-direction and the
$y$-direction are set to $L_x$ and $L_y$, respectively. 
We consider the small static current flowing in the stripe direction 
which is set to the $y$-direction and 
depending only on $x$. 
The injected current generates 
the deviation of the electron density $\delta\rho(x)$ 
from the original value 
in the HF state which also depends only on $x$ 
and gives the energy correction via the 2D Coulomb interaction 
between the deviated charges. 
This energy correction is given by 
\begin{equation}
 \delta E=-\frac{2q^2}{L_x}
  \int^{L_x/2}_{-L_x/2}dxdx'\delta\rho(x)\ln|x-x'|\delta\rho(x'). 
  \label{eq:energy_correction}
\end{equation} 
The deviated charges generates a potential $a_0(x)$. 
The current distribution $\langle j_y(x) \rangle$ and 
$\delta\rho(x)$ are related with $a_0(x)$ as 
$\langle j_y(x)\rangle_a=-\sigma^{(\nu)}_{xy}\partial_x a_0(x)$ and 
$(-e)\delta\rho(x)=2\pi\epsilon \gamma \partial^2_x a_0(x)$ 
in the long wave length limit. 
Here 
$\gamma=(1+\beta \sqrt{B}/\nu)\sigma^{(\nu)}_{xy}/2\pi\epsilon\omega_c$ 
($\sigma^{(\nu)}_{xy}=e^2\nu/2\pi$ is a Hall conductance) 
and 
$\beta$ is zero for the striped Hall state and finite for the ACDW state 
\cite{tsuda}. 
In the case of the integer quantum Hall state, $\beta$ becomes zero. 
The values of $\beta$ for the ACDW state are 
$0.472$ for $l=2$ and $0.581$ for $l=3$ in units of $({\rm tesla})^{-1/2}$. 
The origin of the finite $\beta$ for the ACDW state is 
the density modulation along stripes. 
When the current flows in the stripe direction, 
the deviation of the electron density occurs perpendicularly to the stripe direction.  
The Fermi surface of the striped Hall state has the inter-LL energy
gap in the perpendicular direction to the stripes so that 
$\beta$ becomes zero as in the case of the integer quantum Hall state. 
On the other hand, the ACDW state has the intra-LL energy gap in
addition to the inter-LL energy gap. 
This intra-LL effect gives additional corrections. 
Our study has shown that $\beta$ becomes larger 
as the magnitude of the intra-LL energy gap becomes smaller. 
Since the energy gap of the ACDW state is caused by 
the density modulation along stripes, 
the finite $\beta$ is 
the result of the density
modulation along stripes. 
$a_0(x)$ is approximately given by 
\begin{equation}
 a_0(x)=\alpha\ln\left|\frac{x-L_x/2}{x+L_x/2}\right|\qquad
  {\rm for}\quad |x|\le \frac{L_x}{2}-\gamma,
  \label{eq:a_0}
\end{equation}
with a linear extrapolation of $a_0$ to $\pm IR_H/2$ 
in the interval within $\gamma$ from the edge, 
where $\alpha=IR_H/2(1+\ln(L_x/\gamma))$, and 
$R_H=1/\sigma^{(\nu)}_{xy}$ 
is the Hall resistivity \cite{tsuda}. 
Note that $\gamma$ has the dimension of length and 
is very small for the magnetic fields of order several tesla,  
e.g., if $\epsilon=13\epsilon_0$ and $m=0.067m_e$ (these are parameters in
GaAs), 
then $\gamma$ is of order $10^{-8}$ m. 
For the integer quantum Hall states with $\beta=0$, 
the potential distribution (\ref{eq:a_0}) becomes the same form as 
that obtained by MacDonald et al. \cite{Mac1} and other authors
\cite{Mac2,Mac3,Mac4}.  
The charge distribution obtained from Eq. (\ref{eq:a_0}) 
reveals the accumulation of charges around the both edges
with the opposite sign. 
Substituting Eq. (\ref{eq:a_0}) into
Eq. (\ref{eq:energy_correction}) and performing the $x$-integration, 
we obtain the final result as 
\begin{equation}
 \delta E[I]=C\times I^2 \left(\frac{q^2}{l_{\rm B}}\right),\quad 
 C=\frac{\pi\epsilon}{L_x(\sigma_{xy}^{(\nu)})^2}\times
  \frac{\ln(2/b)-1}{(\ln(2/b)+1)^2}.
  \label{eq:energy_correction2}
\end{equation}
where $b$ is a dimensionless constant given by $b=\gamma/(L_x/2)(\ll 1)$. 

\begin{figure}
\includegraphics[width=7cm]{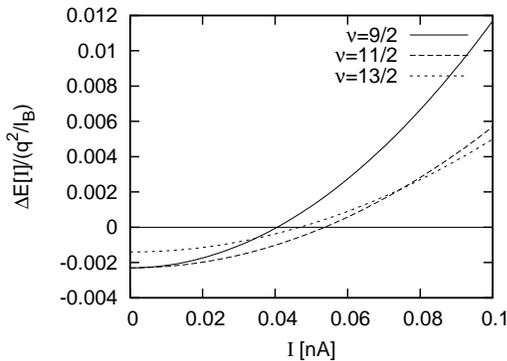}
\caption{\label{fig:ene_diff}Energy differences $\Delta E[I]$
between the striped Hall state 
and the ACDW state. The results at 
$\nu=9/2, 11/2,$ and $13/2$ are shown. 
When $\Delta E[I]$ is positive, 
the striped Hall state has lower energy.}
\end{figure}

The energy correction (\ref{eq:energy_correction2}) depends on the
magnitude of the total current, the filling factor, 
and experimental parameters. 
Since the actual filling factor includes the spin degree of freedom, 
we use $\nu_{\rm ex}= 2l+\nu^\ast$ for lower spin 
bands and $\nu_{\rm ex}= (2l+1)+\nu^\ast$ for upper spin bands 
instead of $\nu$. 
We use $\epsilon=13\epsilon_0$, $m=0.067m_e$, 
$n_e=2.67\times 10^{15}$ m, and $L_x=5\times 10^{-3}$ m ( referred to as
{\it Lilly's parameters}). 
These parameters have been used in the experiment by
M. P. Lilly et al.\cite{stripe_ex1}. 
Substituting Lilly's parameters into Eq. (\ref{eq:energy_correction2}), 
\begin{table}
\caption{Values of the coefficient $C$ in units of ${\rm nA}^{-2}$ and the
 critical current $I_C$ in units of nA. 
}
\label{table:C}
\begin{ruledtabular}
\begin{tabular}{cccc}
$\nu_{\rm ex}$ & $C_{\rm stripe}$ & $C_{\rm ACDW}$   & $I_c$\\
\tableline
9/2  & 144.7  & 146.1   & 0.040 \\
11/2 & 109.9  & 110.7   & 0.053 \\
13/2 & 87.44  & 88.08   & 0.047 \\
\end{tabular}
\end{ruledtabular}
\end{table}
the values of the coefficient $C$ are given as shown in Table \ref{table:C}. 
Including the energy correction due to current, 
the energy difference between the striped Hall state and the ACDW state
is given by (Fig. \ref{fig:ene_diff})
$\Delta E[I]=-\Delta E_0 + (C_{\rm ACDW}-C_{\rm stripe})(q^2/l_{\rm B}) \times (I[{\rm nA}])^2.$ 
The signs of $\Delta E[I]$ change 
at the critical values of current $I_c$. 
The critical values are about $0.04$-$0.05$ nA (Table \ref{table:C}). 
The current used in the experiments \cite{stripe_ex1,stripe_ex2} 
is above $1$ nA and is much larger than the critical value. 
Hence the striped Hall state becomes the lower energy state and 
should be realized in the experiments. 

In summary, 
we have investigated the effect of the injected current  
on the striped Hall state and the ACDW state. 
The injected current flowing in the stripe direction has been
considered. 
It is found that the charge accumulation occurs 
around the both edges with the opposite sign  
and the accumulated charges give energy corrections via the Coulomb
interaction. 
The energies of the two HF states including the 
energy corrections due to current have been compared. 
In the system with no injected current, 
the ACDW state has slightly lower energy than the striped
Hall state. 
We found that the energy of the ACDW state increases faster than 
that of the striped Hall state as the injected current increases. 
The naive expectation suggested from the density modulation along
stripes has been verified. 
Hence the striped Hall state becomes the lower energy state 
when the current exceeds a critical value. 
The critical value is estimated as about $0.04$-$0.05$ nA, 
which is much smaller than the current used in the experiments.
Our results suggest that 
the striped Hall state is realized in
the experiment 
rather than the ACDW state 
and the ACDW state is realized if the experiment 
is done with a current smaller than the critical value. 

\begin{acknowledgments}
This work was partially supported by the special Grant-in-Aid for
Promotion of Education and Science in Hokkaido University, 
a Grant-in-Aid for Scientific Research on Priority Area 
(Dynamics of Superstrings and Field Theories, Grant No. 13135201) 
and (Progress in Elementary Particle Physics of the 21st Century through
Discoveries of Higgs Boson and Supersymmetry, Grant No. 16081201), 
provided by the Ministry of Education, Culture, Sports, Science, and 
Technology, Japan. 
\end{acknowledgments}



\begin{references}
\bibitem{stripe_ex1} M. P. Lilly, K. B. Cooper, J. P. Eisenstein, 
L. N. Pfeiffer, and K. W. West, Phys. Rev. Lett. {\bf 82}, 394 (1999). 

\bibitem{stripe_ex2}
R. R. Du, D. C. Tsui, H. L. Stormer, L. N. Pfeiffer, 
K. W. Baldwin, and K. W. West, Solid State Commun. {\bf 109}, 
389 (1999). 

\bibitem{stripe0} A. A. Koulakov, M. M. Fogler, and B. I. Shklovskii, 
Phys. Rev. Lett. {\bf 76}, 499 (1996); 
M. M. Fogler, A. A. Koulakov, 
and B. I. Shklovskii, Phys. Rev. B {\bf 54}, 1853 (1996). 

\bibitem{stripe02} R. Moessner and J. T. Chalker, 
Phys. Rev. B {\bf 54}, 5006 (1996).

\bibitem{stripe03} E. Fradkin and S. A. Kivelson, Phys. Rev. B {\bf 59}, 
8065 (1999). 

\bibitem{stripe05} E. H. Rezayi, F. D. M. Haldane, and K. Yang, 
Phys. Rev. Lett. {\bf 83}, 1219 (1999). 

\bibitem{stripe01} M. M. Fogler and V. M. Vinokur, Phys. Rev. Lett. 
{\bf 84}, 5828 (2000). 

\bibitem{stripe06} A. H. MacDonald and M. P. A. Fisher, 
Phys. Rev. B {\bf 61}, 5724 (2000). 

\bibitem{ACDW1} R. C\^{o}t\'{e} and H. A. Fertig, Phys. Rev. B {\bf 62}, 
1993 (2000). 

\bibitem{stripe04} V. Oganesyan, S. A. Kivelson, and E. Fradkin, 
Phys. Rev. B {\bf 64}, 195109 (2001). 

\bibitem{stripe07} C. Wexler and A. T. Dorsey, Phys. Rev. B {\bf 64}, 
115312 (2001). 

\bibitem{stripe08} A. Lopatnikova, S. H. Simon, B. I. Halperin, 
and X.-G. Wen, Phys. Rev. B {\bf 64}, 155301 (2001). 

\bibitem{stripe09} N. Shibata and D. Yoshioka, Phys. Rev. Lett. 
{\bf 86}, 5755 (2001). 

\bibitem{stripe10} T. Aoyama, K. Ishikawa, Y. Ishizuka, and N. Maeda, 
Phys. Rev. B {\bf 66}, 155319 (2002); ibid. {\bf 70}, 035314 (2004). 

\bibitem{stripe1} K. Ishikawa, N. Maeda, and T. Ochiai,
 Phys. Rev. Lett. {\bf 82}, 4292 (1999). 

\bibitem{stripe2}
N. Maeda, Phys. Rev. B {\bf 61}, 4766 (2000). 

\bibitem{vNL2}
N. Imai, K. Ishikawa, T. Matsuyama, and I. Tanaka, Phys. Rev. B {\bf 42}, 
10610 (1990). 

\bibitem{vNL3}
K. Ishikawa, N. Maeda, T. Ochiai, and H. Suzuki, Physica E (Amsterdam) {\bf 4E}, 37 (1999). 

\bibitem{ACDW0} D. Yoshioka and H. Fukuyama, 
J. Phys. Soc. Jpn. {\bf 47}, 394 (1979); 
D. Yoshioka and P. A. Lee, Phys. Rev. B {\bf 27}, 
4986 (1983). 

\bibitem{tsuda} K. Tsuda, N. Maeda, and K. Ishikawa, 
cond-mat/0702326. 

\bibitem{Mac1} A. H. MacDonald, T. M. Rice, and W. F. Brinkman, 
Phys. Rev. B {\bf 28}, 3648 (1983). 

\bibitem{Mac2} D. J. Thouless, J. Phys. C {\bf 18}, 6211 (1985). 

\bibitem{Mac3} P. F. Fontein, J. A. Kleinen, P. Hendriks, F. A. P. Blom, 
J. H. Wolter, H. G. M. Lochs, F. A. J. M. Driessen, L. J. Giling, 
and C. W. J. Beenakker, Phys. Rev. B {\bf 43}, 12090 (1991). 

\bibitem{Mac4} C. W. J. Beenakker and H. van Houten, 
in {\it Solid State Phys}, edited by H. Ehrenreich and D. Turnbull 
(Academic, New York, 1992), Vol. 44, pp. 177-181. 


 
\end{references}
\end{document}